\documentclass[debug,overfull]{epl}

\usepackage{graphicx}
\usepackage{dcolumn}
\usepackage{bm}
\usepackage{psfrag}
\usepackage{epsfig}
\usepackage{amsmath}
\usepackage{amssymb}
\usepackage{color}
\usepackage{fancyheadings}
\usepackage{mathbbol}

\newcommand{\bra}{\left\langle}
\newcommand{\ve}{\vert}
\newcommand{\ket}{\right\rangle}

\title{Measurable Consequences of the Local Breakdown of the Concept of Temperature}

\shorttitle{Consequences of Local Breakdown of Temperature}

\author{M. Hartmann\inst{1,2} \and G. Mahler \inst{2}}
\institute{
  \inst{1} Institute of Thechnical Physics, DLR Stuttgart -  Pfaffenwaldring 38-40, D-70569 Stuttgart,
Germany\\
  \inst{2} Institute of Theoretical Physics I, University of Stuttgart - Pfaffenwaldring 57,
D-70550 Stuttgart, Germany\\
hartmann@theo1.physik.uni-stuttgart.de}

\pacs{05.30.-d}{First pacs description}
\pacs{65.80.+n}{Second pacs description}
\pacs{03.67.Mn}{Third pacs description}

\begin{document}

\maketitle

\begin{abstract}
Local temperature defined by a local canonical state of the respective subsystem, does 
not always exist in quantum many body systems. Here, we give some examples of how this 
breakdown of the temperature concept on small length scales might be observed in 
experiments: Measurements of magnetic properties of an anti-ferromagnetic 
spin-1 chain. We show that those magnetic properties are in fact strictly local. 
As a consequence their measurement reveals whether the local (reduced) state can be thermal.
If it is, a temperature may be associated to the measurement results, while this would lead to
inconsistencies otherwise.
\end{abstract}

%
With the advent of nanotechnology, the applicability of thermodynamics on small length
scales has become an important subject of current research. There has been substantial
progress in the fabrication and processing of nanostructures, e.g. carbon nanotubes
\cite{Dresselhaus2001}, and measurements of thermal properties on the corresponding
scales are becoming possible \cite{Gao2002}.
However, despite this amazing experimental advances, the theoretical foundation of
thermodynamics on small scales remains unclear \cite{Cahill2003,Allahverdyan2002}.
First attempts to generalize thermodynamics such that it would apply to small systems
date back to the sixties and are receiving increasing
attention today \cite{Hill2001c,Rajagopal2004}.

Recently \cite{Hartmann2003,Hartmann2003b,Hartmann2004a}, Hartmann, Mahler and Hess
gave a first quantitative estimate of the length scales, below which the standard
concept of temperature ceases to exist. They considered a large chain of particles
with nearest neighbor interactions, which was assumed to be in a thermal, i.e.
canonical, state. In this scenario, subgroups of $N$ adjacent particles are in a
canonical state if $N$ is larger than a threshold value $N_{\text{min}}$ which depends
on the global temperature, $N_{\text{min}} = N_{\text{min}}(T)$. The reduced density matrix
of a subgroup deviates from the canonical form if $N < N_{\text{min}}$.
The temperature dependence of $N_{\text{min}}$ is as follows:
The lower the temperature, the larger $N_{\text{min}}$, i.e. the larger the groups need
to be. In other words, for each group size $N$, there is a threshold temperature above
which the groups are in a canonical state and below they are not.
The predicted length scales depend on the definition of temperature used.
In \cite{Hartmann2003b} and \cite{Hartmann2004a}, local temperature has been defined to
exist if the respective part of the system is in a canonical state. 

Why should we care about the non-existence of local temperature?
There are at least three situations for which this possibility needs special
attention: One obvious scenario refers to the limit of spatial resolution on 
which a temperature profile could be defined. However a spatially
varying temperature calls for non-equilibrium - a complication which we will exclude
here.
A second application deals with partitions on the nanoscale:
If a modular system in thermal equilibrium is partitioned into two pieces, say, the
two pieces need no longer be in a canonical state, let alone have the same local temperature.

Temperature is always measured indirectly via observables, which, in quantum mechanics,
are represented by hermitian operators. Usually, one is interested in measuring the
temperature of a system in a stationary state. The chosen observable should therefore be
a conserved quantity, i.e. its operator should commute with the Hamiltonian of the
system.

A conventional technique, e.g., is to bring the piece of matter, the temperature
$T$ of which is to be measured, in thermal contact with a box (volume $V$) of an ideal
gas (number of particles $n$) and to measure the
pressure $p$ of the gas, which is related to its temperature by $n \, k_B \, T = p \, V$
($k_B$ is Boltzmann's constant).
Since the gas is in thermal equilibrium with the considered piece of matter, both
substances have the same temperature. A measurement of $p$ for constant $V$ allows to
infer the global temperature $T$ of the piece of matter.

Such a thermometer functions as long as it does not significantly perturb the measured object
(weak coupling), irrespective of whether its coupling to the object is local or not.
One might thus doubt whether a local temperature can be measured at all.
System bath models show that the system always relaxes to a canonical state with the
global temperature of the bath, no matter how localized the system bath interaction might be
\cite{Weiss1999}.
However, there are examples, where a local application of Boltzmann-Gibbs thermostatistics is known to
fail due to strong correlations \cite{Netz,Garcia-Morales}.

Here, we consider observables of the object itself, which can be used to measure local temperatures
$T_{\text{loc}}$, i.e. temperatures of subsystems, provided the subsystems are in a canonical state.
In turn, if the respective subsystems are not in a canonical state, this fact
should modify the measurement results for those observables.
For systems composed of weakly interacting subsystems, which are in a thermal state,
the total state factors with respect to the subsystems and the subsystems themselves are also in
thermal states, i.e. local temperature exists. For strongly interacting subsystems, however,
local temperatures cease to exist due to correlations between subsystems.
Note that, if the system reached its thermal state via interaction with a bath, the coupling to the
latter must be weak, otherwise the state of the entire system can, in general, not be thermal
\cite{Allahverdyan2002,Weiss1999}.

Pertinent systems, for which such effects can easily by studied, are magnetic materials.
As we will demonstrate below, properties of single spins can be infered from measurements
of even macroscopic magnetic observables.
These materials thus allow to study the existence of temperature,
as defined by the existence of a canonical state,
on the most local scale possible, i.e. for single spins.

As our model, we consider a homogeneous chain of spin-1 particles interacting with their nearest
neighbors. For the interactions, we assume a Heisenberg model. The Hamiltonian of this
system reads \cite{vanVleck1945}:
\begin{equation} \label{eq:1}
H = B \, \sum_{j=1}^n \sigma_j^z +
J \, \sum_{j=1}^n \vec{\sigma}_j \cdot \vec{\sigma}_{j+1} ,
\end{equation}
where $\vec{\sigma}_j = (\sigma_j^x, \sigma_j^y, \sigma_j^z)$ and
$\vec{\sigma}_j \cdot \vec{\sigma}_{k} =
\sigma_j^x \sigma_k^x + \sigma_j^y \sigma_k^y + \sigma_j^z \sigma_k^z$.
$\sigma_j^x$, $\sigma_j^y$ and $\sigma_j^z$ are the spin-$1$ matrices.
$B$ is an applied magnetic field, $J$ the coupling and $n$ the number of spins.
The coupling $J$ is taken to be positive, $J > 0$. The spins thus tend to align
anti-parallel and the material is anti-ferromagnetic.
The local Hamiltonian of subsystem $j$ is $H_j = B \, \sigma_j^z$.
The system has periodic boundary conditions and is thus
translation invariant.

We assume that the system (\ref{eq:1}) is in a thermal state,
\begin{equation} \label{eq:6}
\rho = \frac{1}{Z} \, \exp \left(- \, \frac{H}{k_B \, T} \right) ,
\end{equation}
where $T$ is the temperature and $Z$ the partition sum. The interaction of this system to
a possible heat bath acting as a thermostat must be weak \cite{Allahverdyan2002,Weiss1999}. 
Note that the temperature $T$ is a global property of the system since it is defined in
the eigenbasis. The eigenstates do not factorize with respect to single spins and
therefore have non-local properties.

The reduced density operator of subsystem $j$ may be represented in the eigenbasis of the
respective subsystem Hamiltonian, $H_j$; let its diagonal matrix elements be $p_{\alpha}$.
It is convenient to introduce a spectral temperature
$T_{\text{spec}}$, which would coincide with $T_{\text{loc}}$ if the local state was canonical
but which formally exists for any state:
\begin{equation}
\label{specdef}
\frac{1}{T_{\text{spec}}} \equiv  - k_B \, \sum_{{\alpha} > 0}
\frac{p_{\alpha}}{1 - p_0} \: \frac{\ln (p_{{\alpha}}) - \ln (p_{0})}{E_{{\alpha}} - E_0} .
\end{equation}
Here the $E_{\alpha}$ are the spectrum of the isolated subsystem;
$E_0$ denotes the ground state. The factor $\left(1 - p_0\right)^{-1}$
is the normalization.
The local state may be characterized by
$T_{\text{spec}}$ and a parameter $\Delta$ describing mean relative square
deviations of the occupation probabilities $p_{\alpha}$ from those of a canonical state,
$p_{\alpha}^c$, with $T_{\text{loc}} = T_{\text{spec}}$,
\begin{equation}
\label{deltadef}
\Delta^2 \equiv \sum_{\alpha} \, p_{\alpha} \:
\left( \frac{p_{\alpha} - p_{\alpha}^c}{p_{\alpha}} \right)^2 ,
\end{equation}
where $p_{\alpha}^c = \exp \left(- E_{\alpha} / k_B \, T_{\text{spec}}\right) /
\sum_{\alpha} \exp \left(- E_{\alpha} / k_B \, T_{\text{spec}}\right)$. Note that $T_{\text{spec}}$
and $\Delta$ depend on the global temperature $T$ {\bf and} the type and strength of
subsystem interactions. $T_{\text{spec}}$ and $\Delta$ do, of course, not fully characterize the local
state (this would require 8 real numbers), they merely classify it in a pertinent way.

\begin{figure}[t]
\centering
\psfrag{0.01}{\hspace{-0.05cm} \small \raisebox{-0.0cm}{$0.01$}}
\psfrag{0.02}{\hspace{-0.05cm} \small \raisebox{-0.0cm}{$0.02$}}
\psfrag{0.03}{\hspace{-0.05cm} \small \raisebox{-0.0cm}{$0.03$}}
\psfrag{0.04}{\hspace{-0.05cm} \small \raisebox{-0.0cm}{$0.04$}}
\psfrag{0.1}{\small \raisebox{-0.0cm}{$0$}}
\psfrag{4.1}{\small \raisebox{-0.0cm}{$4$}}
\psfrag{8.1}{\small \raisebox{-0.0cm}{$8$}}
\psfrag{12.1}{\small \raisebox{-0.0cm}{$12$}}
\psfrag{16.1}{\small \raisebox{-0.0cm}{$16$}}
\psfrag{0}{\small \raisebox{-0.1cm}{$0$}}
\psfrag{2}{\small \raisebox{-0.1cm}{$2$}}
\psfrag{4}{\small \raisebox{-0.1cm}{$4$}}
\psfrag{6}{\small \raisebox{-0.1cm}{$6$}}
\psfrag{8}{\small \raisebox{-0.1cm}{$8$}}
\psfrag{d}{\hspace{-0.2cm}\raisebox{-0.0cm}{$\Delta$}}
\psfrag{s}{\hspace{-1.0cm}\raisebox{0.05cm}{$T / B$ \hspace{0.1cm} ,
\hspace{0.1cm}$T_{\text{spec}} / B$}}
\psfrag{T}{\hspace{-0.3cm}\raisebox{-0.2cm}{$\: T / B$}}
\onefigure[width=12cm]{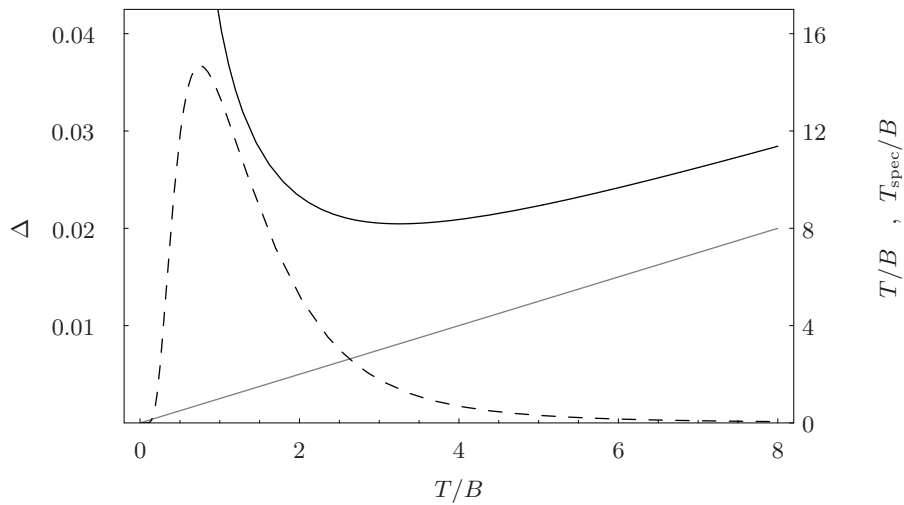}
\caption{\label{fig:2} $T_{\text{spec}}$ (solid line),  $T$ (gray line)
and $\Delta$ (dashed line) as a function of
temperature $T$ for spin-$1$ chain of 4 particles.
$T_{\text{spec}}$ and $T$ are given in units of $B$ and $J = 2 \times B$.}
\end{figure}

Figure \ref{fig:2} shows the spectral temperature $T_{\text{spec}}$, the global
temperature $T$ and the deviations $\Delta$ of the local state from a canonical state with
$T_{\text{spec}}$ as a function of $T$ for a spin-$1$ chain of 4 particles
with the Hamiltonian (\ref{eq:1}).
While the deviations $\Delta$ are small at high
$T$, they become larger for low $T$, where the spectral temperature $T_{\text{spec}}$
starts to rise again as $T$ is lowered further.
For $T = 0$, $\Delta$ vanishes, since a completely mixed state corresponds to a canonical
one with $T_{\text{spec}} \rightarrow \infty$.
A signature of these local deviations from a canonical state ($\Delta \not= 0$)
can be measured.
%

As an example of such an experiment, we will now consider two different magnetic
observables of a spin-$1$ system with the Hamiltonian (\ref{eq:1}).
The first observable is the magnetization in the direction of the applied field,
$m_z$, which we define to be the total magnetic moment per particle: 
\begin{equation} \label{eq:2}
m_z \equiv \frac{1}{n} \bra \sum_{j=1}^n \sigma_j^z \ket ,
\end{equation}
where $\bra \mathcal{O} \ket$ is the expectation value of the operator $\mathcal{O}$,
i.e. $\bra \mathcal{O} \ket = \text{Tr}(\rho \, \mathcal{O})$.
In the translation invariant state $\rho$, the reduced
density matrices of all individual spins are equal, and the magnetization (\ref{eq:2})
can be written as
\begin{equation} \label{eq:3}
m_z = \bra \sigma_k^z \ket ,
\end{equation}
for any $k = 1, 2, \dots, n$.
The magnetization, although defined macroscopically, is thus actually a
property of a single spin, i.e. a strictly local property.

As our second observable we choose the occupation probability, $p$, of the $s_z = 0$
level (averaged over all spins),
\begin{equation} \label{eq:5}
p \equiv \frac{1}{n} \bra \sum_{j=1}^n \left\ve 0_j \ket \bra 0_j \right\ve \ket
 = \bra \left\ve 0_k \ket \bra 0_k \right\ve \ket ,
\end{equation}
where the second equality holds for the same reasons as for $m_z$ for any $k = 1, 2, \dots, n$.
$p$ is thus strictly local, too.

Now, if each single spin was in a canonical state with $\Delta = 0$ and a temperature
$T_{\text{loc}} = T_{\text{spec}}$,
$m_z$ and $p$ would both have to be monotonic functions of $T_{\text{loc}}$.
Consequently, $T_{\text{loc}}$ could, after calibration,
be infered from measurements of $m_z$ or $p$. Note, that
$m_z$ is proportional to the local energy, the average energy of one subsystem.

\begin{figure}
\centering
\psfrag{-0.2}{\small \raisebox{-0.0cm}{$-0.2$}}
\psfrag{-0.4}{\small \raisebox{-0.0cm}{$-0.4$}}
\psfrag{-0.6}{\small \raisebox{-0.0cm}{$-0.6$}}
\psfrag{-0.8}{\small \raisebox{-0.0cm}{$-0.8$}}
\psfrag{-1.1}{\small \raisebox{-0.0cm}{$-1.0$}}
\psfrag{0}{\small \raisebox{-0.1cm}{$0$}}
\psfrag{2}{\small \raisebox{-0.1cm}{$2$}}
\psfrag{4}{\small \raisebox{-0.1cm}{$4$}}
\psfrag{6}{\small \raisebox{-0.1cm}{$6$}}
\psfrag{8}{\small \raisebox{-0.1cm}{$8$}}
\psfrag{0.4}{\small \raisebox{-0cm}{$0.4$}}
\psfrag{0.3}{\small \raisebox{-0cm}{$0.3$}}
\psfrag{0.2}{\small \raisebox{-0cm}{$0.2$}}
\psfrag{0.1}{\small \raisebox{-0cm}{$0.1$}}
\psfrag{1.2}{\small \raisebox{-0cm}{$0.0$}}
\psfrag{m}{\hspace{-0.05cm}\raisebox{0.0cm}{$m_z$}}
\psfrag{p}{\hspace{0.05cm}\raisebox{0.0cm}{$p$}}
\psfrag{T}{\hspace{-0.2cm}\raisebox{-0.2cm}{$\: T / B$}}
\onefigure[width=11cm]{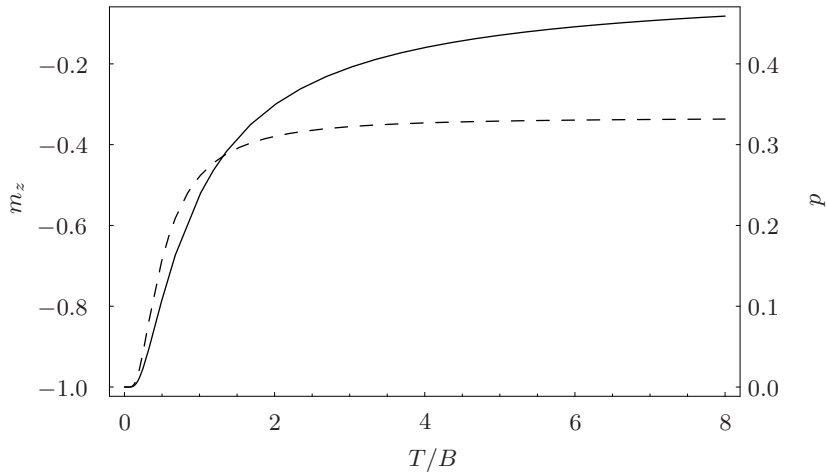}
\caption{\label{fig:1} $m_z$ (solid line) and $p$ (dashed line) as a function of
temperature $T$ for spin-$1$ chain of 4 particles.
$T$ is given in units of $B$ and $J = 0.1 \times B$.}
\end{figure}

\begin{figure}
\centering
\psfrag{-0.05}{\hspace{0.1cm} \small \raisebox{-0.0cm}{$ 0.00$}}
\psfrag{-0.02}{\small \raisebox{-0.0cm}{$ $}}
\psfrag{-0.04}{\small \raisebox{-0.0cm}{$-0.04$}}
\psfrag{-0.06}{\small \raisebox{-0.0cm}{$ $}}
\psfrag{-0.08}{\small \raisebox{-0.0cm}{$-0.08$}}
\psfrag{-0.1}{\small \raisebox{-0.0cm}{$ $}}
\psfrag{-0.12}{\small \raisebox{-0.0cm}{$-0.12$}}
\psfrag{-0.14}{\small \raisebox{-0.0cm}{$ $}}
\psfrag{0}{\small \raisebox{-0.1cm}{$0$}}
\psfrag{2}{\small \raisebox{-0.1cm}{$2$}}
\psfrag{4}{\small \raisebox{-0.1cm}{$4$}}
\psfrag{6}{\small \raisebox{-0.1cm}{$6$}}
\psfrag{8}{\small \raisebox{-0.1cm}{$8$}}
\psfrag{0.332}{\small \raisebox{-0cm}{$0.332$}}
\psfrag{0.334}{\small \raisebox{-0cm}{$ $}}
\psfrag{0.336}{\small \raisebox{-0cm}{$0.336$}}
\psfrag{0.338}{\small \raisebox{-0cm}{$ $}}
\psfrag{0.34}{\small \raisebox{-0cm}{$0.340$}}
\psfrag{0.342}{\small \raisebox{-0cm}{$ $}}
\psfrag{0.344}{\small \raisebox{-0cm}{$0.344$}}
\psfrag{0.346}{\small \raisebox{-0cm}{$ $}}
\psfrag{m}{\hspace{-0.05cm}\raisebox{0.0cm}{$m_z$}}
\psfrag{p}{\hspace{0.05cm}\raisebox{0.0cm}{$p$}}
\psfrag{T}{\hspace{-0.2cm}\raisebox{-0.2cm}{$\: T / B$}}
\onefigure[width=12cm]{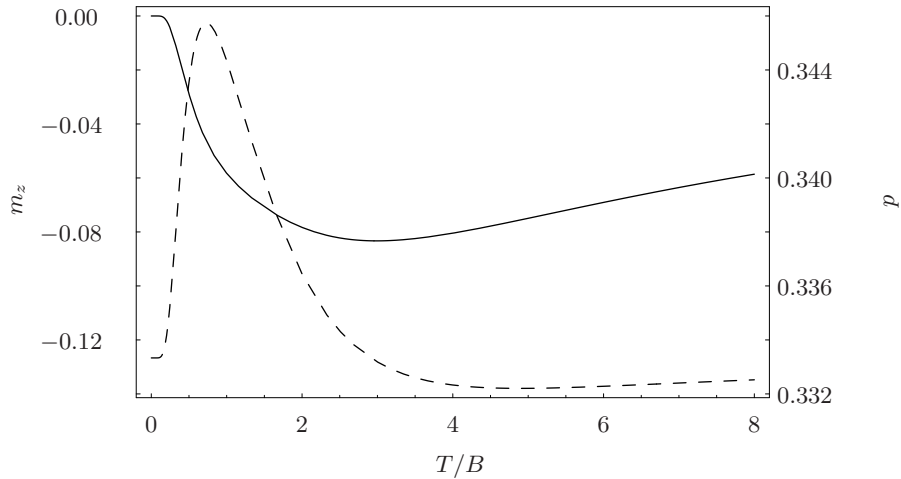}
\caption{\label{fig:3} $m_z$ (solid line) and $p$ (dashed line) as a function of
temperature $T$ for spin-$1$ chain of 4 particles.
$T$ is given in units of $B$ and $J = 2 \times B$.}
\end{figure}

Figure \ref{fig:1}, shows $m_z$ and $p$ as a function
of the global temperature $T$ for
a spin-$1$ chain of 4 particles with the Hamiltonian (\ref{eq:1})
for weak interactions, $J = 0.1 \times B$.
Both quantities are monotonic functions of each other.
%
The situation changes drastically when the spins are strongly coupled.
In this case the concept of temperature breaks down locally due to
correlations of each single spin with its environment.
Figure \ref{fig:3} shows $m_z$ and $p$ as a function of temperature $T$ for
a spin-$1$ chain of 4 particles with the Hamiltonian (\ref{eq:1})
for strong interactions $J = 2 \times B$.
Both quantities are non-monotonic functions of $T$ and therefore no mapping
between $m_z$ and $p$ exists.
%
Note that, although the present plots are done for $4$ spins only, the characteristics do
not change for larger numbers of spins (The exact quantitative values are not needed for our
reasoning). We have studied this with chains of $4$, $5$ and $6$ spins.
In the case of $m_z$, the characteristics have even been already observed in experiments
\cite{Avenel1992}. 

How could a local observer determine whether the system he observes, a single spin,
is in a thermal state and can therefore be characterized by a temperature?
The local observer needs to compare two situations:
In the first situation, the spin is weakly coupled to a larger system, the heat bath.
In this situation, the local observer could measure $m_z$ and $p$ as functions of 
the temperature of the heat bath and would get a result similar to figure \ref{fig:1}.
This result would not be sensitive to the details of the coupling to the
heat bath. The local observer would thus recognize this situation as a particular one
and might term it the ``thermal'' situation.
The second situation is fundamentally different. The spin is now strongly coupled to
its surrounding. If the local observer again measures $m_z$ and $p$ as functions of 
the temperature of the surrounding, he gets the result of figure \ref{fig:3}.

The observer can tell the difference between both situations, even if he has no access to the
temperature $T$ of the surrounding. In the first case he can construct a mapping from
say $m_z$ to $p$ or vice versa, in the second he cannot:
Here the concept of a local temperature breaks down
at least on the level of individual particles, since temperature measurements
via different local observables would contradict each other.

Finally, we address the question of whether the effects described here could be observed
in real experiments. Indeed, pertinent experiments are available and have partly
already been carried out:
A realization of a quasi one dimensional anti-ferromagnetic spin-1 Heisenberg chain is
the compound CsNiCl$_3$
\cite{Kenzelmann2001,Affleck1989,Avenel1992}. Here the coupling is
$J \approx 2.3$~meV. To achieve a detectable modulation of $m_z$ and $p$, the spins
should be significantly polarized for $T > 0$.
Therefore a sufficiently strong applied magnetic field is needed.
For CsNiCl$_3$, a field of roughly $9.8$~Tesla would correspond to $J = 4 \times B$.

The magnetization in an applied field can be measured with high precision with a
SQUID \cite{Lipa1981}. The occupation probability of the $s_z = 0$ states, on the other
hand, is accessible via neutron scattering experiments
\cite{Kenzelmann2002a,Ma1995}.
The differential cross section for neutron scattering of spin-systems is given by
\cite{Lovesey1989}
\begin{equation} \label{eq:8}
\frac{\partial^2 \sigma}{\partial \Omega \, \partial E_f} \sim \left| f(\vec{q})^2 \right|
\frac{\left| \vec{k}_f \right|}{\left| \vec{k}_i \right|} \sum_{a, b}
\left( \delta_{a, b} -
\frac{\vec{q}_a}{\left|\vec{q}_a\right|} \cdot \frac{\vec{q}_b}{\left|\vec{q}_b\right|}
\right) S^{a b} (\vec{q}, \omega) ,
\end{equation}
where $E_i$, $E_f$, $\vec{k}_i$ and $\vec{k}_f$ are the initial and final energies and
momenta of the scattered neutrons. $a, b = x, y, z$, $\omega = E_f - E_i$,
$\vec{q} = \vec{k}_f - \vec{k}_i$ and $f(\vec{q})$ is the magnetic form factor,
which can be found tabulated \cite{Brown1999}. The dynamic structure factor
$S^{a b} (\vec{q}, \omega)$ is the Fourier transform of the spin-spin correlation
function,
\begin{equation}
S^{a b} (\vec{q}, \omega) = \frac{1}{2 \pi} \int dt \sum_{\vec{r}, \vec{r}'}
e^{i \vec{q} (\vec{r} - \vec{r}') - i \omega t}
\bra \sigma^a_{\vec{r}}(0) \, \sigma^b_{\vec{r}'}(t) \ket .
\end{equation}
If we only consider events, where the difference in momentum is along the z-axis,
$\vec{q} = q_z \vec{e}_z$, only $S^{x x}$, $S^{x y}$, $S^{y x}$ and $S^{y y}$
contribute in equation (\ref{eq:8}). Since the applied field $B$ is along the z-axis,
$S^{x y}$ and $S^{y x}$ are zero due to symmetry reasons.
Summing up over all $\vec{q}$ and all $\omega$ and using our knowledge of
$\vec{k}_i$, $\vec{k}_f$ and $f(\vec{q})$, we can obtain information about the quantity
\begin{equation}
\frac{1}{n} \, \sum_{\vec{r}}
\bra \sigma^x_{\vec{r}}(0) \, \sigma^x_{\vec{r}}(0) +
\sigma^y_{\vec{r}}(0) \, \sigma^y_{\vec{r}}(0) \ket = 1 + p
\end{equation}
from the measurement data.
Therefore, $p$ is measurable in neutron scattering experiments.
Such experiments or a combination thereof could thus be used to demonstrate the
non-existence of local temperature.

We note in passing that such ``local'' measurements are also an interesting tool to study local
and non-local features like entanglement of thermal quantum
states \cite{Jordan2003,oConnor2001}.

In summary, we have discussed possible ``local'' temperature measurements via magnetic properties
of the considered material. We have studied a magnetic structure in a
thermal state and analyzed whether its ``local'' temperature could be infered from the
measurement of magnetic quantities.
For a large structure, for which boundary effects can be neglected and thus
a model with periodic boundary conditions is a valid description, the magnetization is
a strictly local property implying that a temperature measurement based on it is strictly
local, too.
Local properties, in general, need not be monotonic functions of the
global temperature and thus cannot serve as temperature indicators.
The global temperature is not accessible by a local measurement.

For networks of spin-1 or higher dimensional subsystems, two local
but different quantities like the magnetization and the occupation probability of the
$s_z = 0$ state, cannot be mapped onto each other.
As a consequence, using these two quantities for temperature measurements would yield
contradictive results.
These measurements thus show, that the local state cannot be canonical and that
therefore a local temperature does not exist for single spins.

The existence of non-thermal properties within a modular system in thermal equilibrium has
so far rarely been recognized.
The popularity of a thermal description and
the notion of temperature is based on the fact that various properties of the system uniquely
scale with temperature. If these mappings do no longer exist, the concept of temperature looses
its meaning. This fact might become relevant in future nanotechnologies, in so far as the behavior of a 
structure can no longer be predicted from a temperature that would characterize it. 

We thank M.\ Henrich, C.\ Kostoglou, M.\ Michel, H.\ Schmidt, M.\ Stollsteimer
and F.\ Tonner for fruitful discussions.
M.\ Hartmann wants to thank Shuichi Wakimoto for comments on the experimental techniques,
Prof. Ulrich Schollw\"ock for comments on some pertinent
experimental materials and, in particular, Prof. Ortwin Hess for discussions
and support, which made this work possible.

%
%

%
%

\end{document}